\documentclass[aps,pre,preprint,unsortedaddress,longbibliography]{revtex4-1}
\usepackage{graphicx}
\usepackage{epstopdf, epsfig}
\usepackage{dcolumn}
\usepackage{bm}
\usepackage{lineno}
\usepackage{xcolor}
\usepackage{algpseudocode} 
\usepackage[]{algorithmicx}
\usepackage{varwidth}
\usepackage{color,soul}

\usepackage{ulem}

\begin{document}
\title{\textcolor{black}{Deformation and breakup dynamics of droplets within a tapered channel}}

\author{Andrea Montessori}
\affiliation{Istituto per le Applicazioni del Calcolo CNR, via dei Taurini 19, 00185, Rome, Italy}

\author{Michele La Rocca }
\affiliation{Dipartimento di Ingegneria, Università degli Studi Roma TRE, via Vito Volterra 62, Rome, 00146, Italy}

\author{Pietro Prestininzi }
\affiliation{Dipartimento di Ingegneria, Università degli Studi Roma TRE, via Vito Volterra 62, Rome, 00146, Italy}

\author{Adriano Tiribocchi}
\affiliation{Istituto per le Applicazioni del Calcolo CNR, via dei Taurini 19, 00185, Rome, Italy}

\author{Sauro Succi}
\affiliation{Center for Life Nanoscience @ La Sapienza, Istituto Italiano di Tecnologia, viale Regina Elena 295, 00161, Rome, Italy}
\affiliation{\textcolor{black}{Istituto per le Applicazioni del Calcolo CNR, via dei Taurini 19, 00185, Rome, Italy}}
\affiliation{Institute for Applied Computational Science, Harvard John A. Paulson School of Engineering and Applied Sciences, Cambridge, MA 02138, United States}
\date{\today}

\begin{abstract}

In this paper we numerically investigate the breakup dynamics of droplets in an emulsion flowing in a tapered microchannel with a narrow constriction.
The mesoscale approach for multicomponent fluids with near contact interactions is shown to capture the deformation and breakup dynamics of droplets interacting within the constriction, in agreement with experimental evidences. In addition, it permits to investigate in detail the hydrodynamic phenomena occurring during the breakup stages.
Finally, a suitable deformation parameter is introduced and analyzed to characterize the  state of deformation of the system by inspecting pairs of interacting droplets flowing in the narrow channel. 
\end{abstract}

\maketitle

\section*{Credit Line}

The following article has been accepted by Physics of Fluids. After it is published, it will be found at 
\url{https://aip.scitation.org/journal/phf}

\section{Introduction}

In recent years droplet-based microfluidics, namely the art of manipulating, controlling and fine-tuning the production of  emulsions characterized by a high degree of structural ordering and monodispersity \cite{marmottant2009microfluidics,raven2009,garstecki2006flowing}, has witnessed a giant leap forward, mostly thanks to the advancements in miniaturization processes which, in turn,  paved the way to a clever design of micron and submicron-sized channels and devices \cite{Sima2021pof,Vagner2021pof,leong2018internal}. These technological boosts sustained a surge of experimental \cite{costantini2015micro,weitz} and theoretical \cite{montessoriprf,montessori2019jetting,tiribocchi2021shear} activities aimed at shedding light on the underlying mechanisms responsible for the emergence of complex dynamical behaviors in densely packed systems evolving in strongly confined geometries. 

Although both experimental and theoretical research have already driven many improvements towards their understanding, the investigation of droplet-based environments poses a steep challenge owing  to i)  the variety of spatial scales in play, ii) the presence of highly non-linear behaviors, often out of reach for  analytical, perturbative approaches and iii) the presence of many local field quantities, whose detailed knowledge is of decisive importance for a clever design of novel microfluidic devices. From this standpoint, an accurate numerical modeling of these complex, many-body, interacting systems stands as a way to inspect in closer details a number of peculiar phenomena occurring at different spacetime scales and to unveil a plethora of complex behaviors otherwise inaccessible neither to theoretical nor to experimental approaches. The present work, which falls precisely within this line of thinking, aims at presenting \textit{ab-initio} hydrodynamic, mesoscale simulations of deformation and breakup phenomena occurring in dense emulsions, namely monodisperse ensembles of interacting droplets immersed in a bulk fluid flowing within a tapered microfluidic channel \cite{bick2019effect,gai2016confinement,gai2016spatiotemporal}.

\textcolor{black}{Indeed, a particularly relevant phenomenon usually compromising the accuracy of the resulting emulsion is the breakup of droplets, especially in dense suspensions where the volume fraction exceeds the close-packing limit \cite{rosenfeld2014break,gai2016confinement,bick2019effect}. 
Experimental results of a single droplet under an external flow support the view that deformation and breakup are determined by a delicate balance between viscous stress and interfacial tension, described by the Capillary number ($Ca$). While the viscous stress tends to deform the drop, the interfacial tension contrasts this effect and holds the drop in a spherical shape. Once the capillary number overcomes a critical value, viscous stress dominates and the breakup becomes inevitable. In concentrated emulsions, in contrast,
the volume fraction of the droplet phase significantly increases, and experiments suggest that breakup arises from droplet-droplet and droplet-wall interactions \cite{gai2016confinement,bick2019effect,khor2017time,gai2017amphiphilic}
However many questions remain open.\\
At variance with the aforementioned experimental works, the goal of the present work is to pinpoint the effect of viscous dissipation over surface tension forces of individual droplets of a concentrated emulsion flowing within a tapered channel. This target is achieved by varying the Capillary number while keeping the Reynolds number constant, in order to decouple the effects due to inertia, surface tension and viscous forces, a task generally difficult (if not impossible) to realize experimentally. Besides confirming the results described in Ref.\cite{gai2016confinement}, our model, being capable to deliver the full hydrodynamic picture, offers a way to unveil highly non-trivial details of the structure of the velocity field observed during the breakup stages. In addition, by tracking the circularity of hundreds of interacting droplets,  we find that the maximum deformation, before a breakup event, is independent of $Ca$ and the number of breakups occurring at the constriction linearly increases with $Ca$.}




%
\section{Method}
In the following we briefly describe the extended mesoscale approach for multicomponent flows with interacting interfaces developed in \cite{montessori2019mesoscale,montessori2019modeling}. 
The multicomponent system, namely an emulsion formed by droplets immersed in a bulk fluid, is modeled via a color-gradient \cite{gunstensen1991,rothman1988immiscible} regularized Lattice Boltzmann  method \cite{montessori2015lattice,latt2006lattice,coreixas2019comprehensive}, which employs two sets of probability distribution functions. Each set evolves via a sequence of streaming and collision steps \cite{kruger2017lattice,succi2018lattice,benzi1992lattice}
\begin{equation} \label{LBM1}
f_i^k(\mathbf{x}+\mathbf{c}_i \Delta t, t+\Delta t)=f_i^k(\mathbf{x},t) + \Omega_i^k (f_i^k(\mathbf{x},t)) + F_i^{rep},
\end{equation}
where $f_i^k(\mathbf{x},t)$ is the $i^{th}$ discrete probability distribution function for the $k^{th}$ component, giving the probability of finding a fluid particle at position $\mathbf{x}$, time $t$ and with discrete velocity $\mathbf{c}_i$. The index $k$ is such that $k=1,2$, while $i$ belongs to the range $0 \le i \le N_{set}$, being $N_{set}$ the dimension of the set of probability distribution functions. This is equal to $8$ for the two-dimensional nine speeds lattice (D2Q9) adopted in this paper.
The time step $\Delta t$ is expressed in lattice units \cite{kruger2017lattice} and set to $1$, a usual choice in the LBM \cite{succi2018lattice}. Finally, $F_i^{rep}$ is a force aimed at upscaling  the repulsive near-contact interactions acting on scales much smaller than the resolved ones \cite{montessori2019mesoscale}.

Once the set of distribution is known at each lattice site, the macroscopic fields of interest, namely the fluid density $\rho_k$, the linear momentum $\rho_k \mathbf{u}_k$ and the pressure $p_K$, can be obtained by computing the relevant statistical moments:
\begin{equation} \label{LBM2}
\rho_k = \sum_{i=0}^{N_{set}} f_i^k(\mathbf{x},t), \\ \rho_k \mathbf{u}_k = \sum_{i=0}^{N_{set}} \mathbf{c}_i f_i^k(\mathbf{x},t), \ p_k=\frac{\rho_k}{3}.
\end{equation}
The total fluid density $\rho$ and the total momentum $\rho \mathbf{u}$ of the mixture can be obtained via the following relations:
\begin{equation} \label{LBM3}
\rho = \sum_{k=1}^{2} \rho_k, \\ \rho \mathbf{u} = \sum_{k=1}^{2} \rho_k \mathbf{u}_k. 
\end{equation}
It is worth recalling that the multicomponent approach employed in this work is based on a variant of the color-gradient LB model. By employing the standard formalism, the  collision operator can be split into three parts \cite{leclaire2012numerical,leclaire2017}:
\begin{equation} \label{LBM4}
\Omega_i^k = (\Omega_i^{k})^{(3)} [(\Omega_i^{k})^{(1)}+(\Omega_i^{k})^{(2)}].
\end{equation}
The term $(\Omega_i^{k})^{(1)}$ is the usual single relaxation time Bhatnagar–Gross–Krook collisional operator \cite{succi2018lattice}
\begin{equation} \label{LBM5}
(\Omega_i^{k})^{(1)}=\frac{f_{i,eq}^k(\mathbf{u},\mathbf{x},t)-f_i^k(\mathbf{x},t)}{\tau},
\end{equation}
where $\tau$ is the effective relaxation time, which depends on the viscosity $\nu$ of the emulsion
\begin{equation} \label{LBM6}
\nu=\left(\frac{\rho_1}{\rho_1+\rho_2}\frac{1}{\nu_1} +\frac{\rho_2}{\rho_1+\rho_2}\frac{1}{\nu_2}\right)^{-1}  , \  \tau= 3 \nu + \frac{1}{2}. 
\end{equation}
Here $\nu_1, \nu_2, \rho_1, \rho_2$ are the kinematic viscosities and the densities of the two fluids in the bulk, respectively. Also, $f_{i, eq}^k(\mathbf{x},t)$ is the $i^{th}$ equilibrium discrete probability distribution function for the $k^{th}$ component, formally derived as a low Mach, second-order expansion of a Maxwellian probability equilibrium distribution \cite{succi2018lattice}
\begin{equation} \label{LBM7}
f_{i, eq}^k(\mathbf{u},\mathbf{x},t)=w_i \rho_k \Bigg(1+3 \mathbf{c}_i \cdot \mathbf{u} + 9 \frac{(\mathbf{c}_i \cdot \mathbf{u})^2}{2} -3  \frac{\mathbf{u} \cdot \mathbf{u}}{2} \Bigg),
\end{equation}
being $w_i$ the weights of the D2Q9 lattice  adopted in this paper \cite{succi2018lattice}.\\
The term $(\Omega_i^{k})^{(2)}$ is the perturbation part, which accounts for the interfacial tension, and reads as
\begin{equation} \label{LBM8}
(\Omega_i^{k})^{(2)}= \frac{A_k}{2} |\nabla{\Theta}|\Bigg( w_i \bigg( \frac{\mathbf{c}_i \cdot \nabla{\Theta}}{|\nabla{\Theta}|} \bigg)^2-B_i \Bigg),
\end{equation}
where $A_k$ ($k=1,2)$ and $B_i$  $(i=0,N_{set})$ are suitable constants defined in \cite{montessori2019mesoscale}. $\Theta$ is a scalar phase field, defined as
\begin{equation} \label{LBM9}
\Theta=\frac{\rho_1-\rho_2}{\rho_1+\rho_2},
\end{equation}
assuming the value $1$ in the fluid component with density $\rho_1$ and the value $-1$ in the fluid component with density $\rho_2$. It is worth observing that the constants $A_1, A_2$ are related to the surface tension $\sigma$ by
\begin{equation} \label{LBM10}
\sigma=\frac{2}{9} \tau (A_1+A_2).
\end{equation}
Finally the third term $(\Omega_i^{k})^{(3)}$ is the recolouring operator which aims at minimizing the mutual diffusion between the fluid components, thus favouring their separation \cite{latva2005diffusion}
\begin{equation} \label{LBM11}
 f_i^1(\mathbf{x},t)  = \frac{\rho_1}{\rho} \sum_{k=1}^2 f_i^{*k}(\mathbf{x},t) + \beta \frac{\rho_1 \rho_2}{\rho^2} \Bigg( \frac{\nabla \Theta \cdot \mathbf{c}_i}{|\nabla \Theta|} \Bigg) \sum_{k=1}^2  f_{i, eq}^k(0,\mathbf{x},t),
\end{equation}
\begin{equation} \label{LBM12}
 f_i^2(\mathbf{x},t)  =\frac{\rho_2}{\rho} \sum_{k=1}^2 f_i^{*k}(\mathbf{x},t) - \beta \frac{\rho_1 \rho_2}{\rho^2} \Bigg( \frac{\nabla \Theta \cdot \mathbf{c}_i}{|\nabla \Theta|} \Bigg) \sum_{k=1}^2  f_{i, eq}^k(0,\mathbf{x},t).
\end{equation}
Here $f_i^{*k}(\mathbf{x},t) \ (i=0,N_{set}; \ k=1,2)$ are the post-collision and post-perturbation probability distribution functions, while the coefficient $\beta$ is a free parameter which tunes the width of the interface.\\
By performing a multiscale Chapman–Enskog expansion, it can be shown that the hydrodynamic limit of Eq.(\ref{LBM1}), in the low frequency-long wavelength and low Mach number limit, is a set of conservation laws for mass and linear momentum
\begin{equation} \label{LBM13}
\frac{\partial \rho}{\partial t} + \nabla \cdot (\rho \mathbf{u})=0
\end{equation}
\begin{equation} \label{LBM14}
\frac{\partial (\rho \mathbf{u}) }{\partial t} + \nabla \cdot (\rho \mathbf{u} \mathbf{u})=- \nabla p + \nabla \cdot ( \rho \nu (\nabla (\mathbf{u} + \mathbf{u}^T )) + (\nabla \sigma - \sigma (\nabla \cdot \mathbf{n}) \mathbf{n} - A_h \mathbf{n}) \delta_I,
\end{equation}
where $p=\sum_{k=1}^2 p_k$ is the total pressure. The last term at the right hand side of Eq.(\ref{LBM14}) codes for the effects of the surface tension $\sigma$ and for the additional, short-range, repulsive force at the interface between the two fluid components. In Eq.(\ref{LBM14}), $\delta_I$ is an index function, defined as $\delta_I=|\nabla \Theta|/2$, which localizes the force on the interface, while $A_h$ is a parameter controlling the strength of the repulsive force oriented as the local normal $\mathbf{n}$ to the interface \cite{montessori2019mesoscale}.
\section{Results}

In the following, after a brief description of the numerical setup employed, we discuss in detail the deformation and breakup phenomena by tracking the evolution of the droplets flowing within the narrow channel.

\subsection{Numerical setup}

\begin{figure}
\centerline{
\includegraphics[scale=0.85]{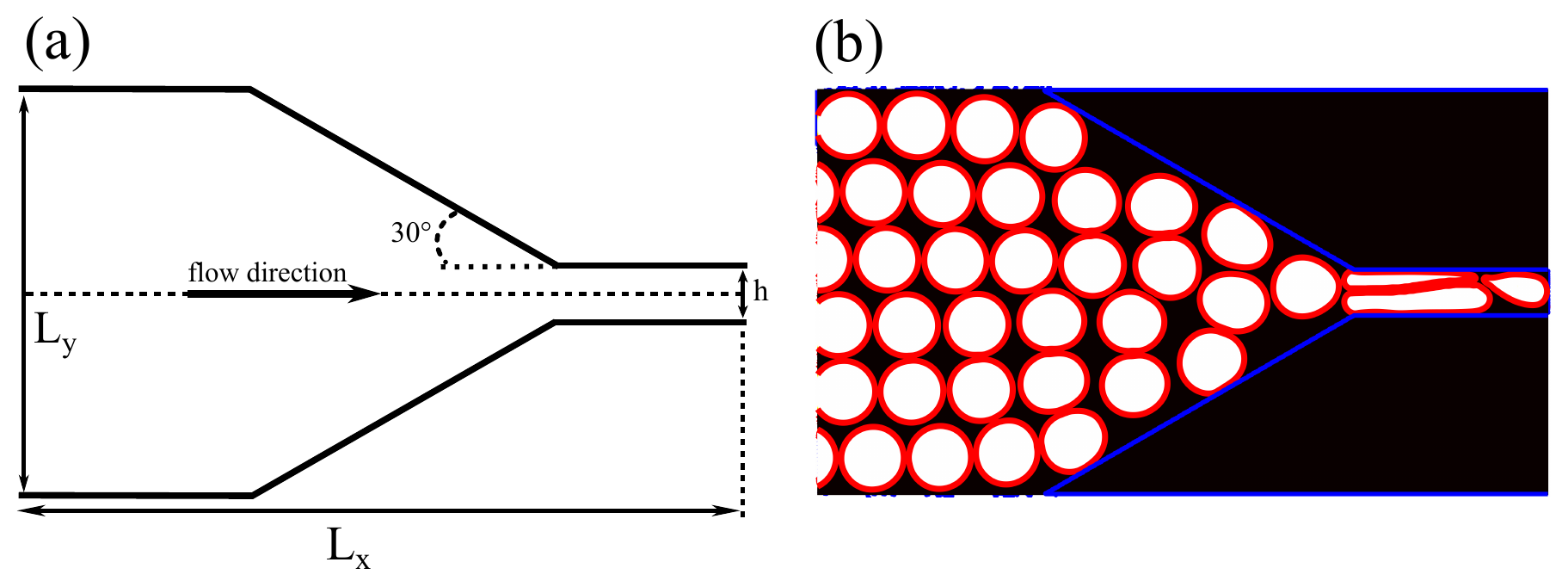}}
\caption{Panel (a): Sketch of the microchannel geometry employed in this work. Panel (b): Snapshot of the droplets' field flowing within the tapered channel. The droplets are continuously injected at the inlet via an internal periodic boundary condition, as proposed in \cite{montessori2021mesoscale,montessori2021wet}.}
\label{FIGURA1}
\end{figure}

The geometrical setup, conveniently sketched in fig.\ref{FIGURA1}(a), consists of a tapered geometry with an outlet narrow channel positioned at its end.  The height of the domain is $L_y=416lu$ (lattice units), its length is $L_x=750lu$, while the height of the constriction is $h\sim0.75d=45lu$, being $d$ the diameter of the droplet. Finally, the angle of the tapered geometry is set to $30^\circ$ as in \cite{gai2016confinement,bick2019effect,gai2016spatiotemporal,rosenfeld2014break}.  
The emulsion, flowing within the microchannel and continuously injected via an internal periodic boundary condition as in \cite{montessori2021wet,montessori2021mesoscale}, is made of a set of monodisperse droplets of radius $d/2$, as shown fig.\ref{FIGURA1}(b). 
\textcolor{black}{The relevant parameters employed in the simulations are the following: kinematic viscosity $\nu=0.023$, droplets' reference velocity $U=0.018$, droplet's diameter $d=60$, while the  surface tension ranges between $\sigma=0.006\div0.04$.}\\
It is worth noting that the viscosity $\nu$ (taken equal in both the dispersed and the bulk fluid) and the inlet carrier velocity $U$ were set to constant values in all simulations, while the surface tension was varied within a range of an order of magnitude. \textcolor{black}{ This particular choice permits to vary the Capillary number ($Ca=\rho U \nu/\sigma=0.011\div0.072$) independently of the Reynolds number ($Re=U h/\nu$)}, which instead was kept fixed in all simulations.
By doing so, we isolated the effects of the variations of the elongational viscous forces versus the surface tension forces, regardless of the inertia. 

\subsection{Deformation dynamics of interacting droplets within a constriction}

\begin{figure}
\centerline{
\includegraphics[scale=0.9]{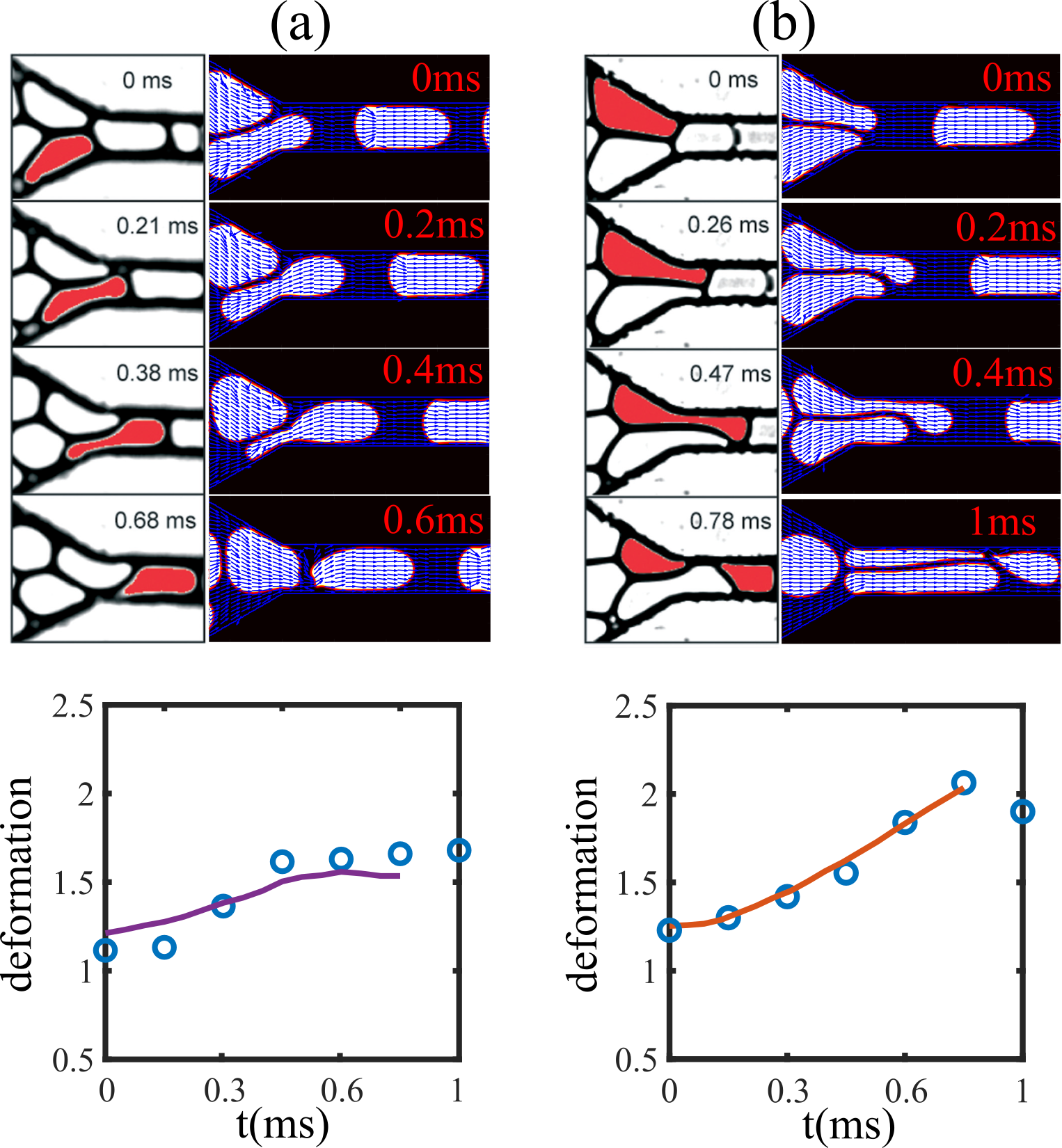}}
\caption{Droplets' deformation and breakup in the narrow constriction at $Ca=0.043$. Left columns report experimental results \cite{gai2016confinement}, while the right columns show the simulations. Two possible outcomes may occur once the droplets concur to enter the narrow channel. In the first one (panel(a)), the droplet undergoes a fast expansion which favors its escape towards the outlet of the constriction. In the second one (panel (b)), the two droplets stretch under the effect of the viscous forces acting at the interface level until one breaks. \textcolor{black}{Bottom panels. Comparison between experiments\cite{gai2016confinement} (solid line) and simulations (open circles) of the deformation history.}}
\label{FIGURA1BIS}
\end{figure}

Firstly, we start with a  brief description of the salient dynamical features emerging during the interaction of a pair of flowing droplets. 
With reference to fig.\ref{FIGURA1BIS} (showing two sequences of droplets interacting within the microchannel), once the droplets approach the constriction,  they start to deform under the effect of the confinement, which is caused by i) the walls of the constriction and ii) the presence of the neighboring droplet. As evidenced in  fig.\ref{FIGURA1BIS}, two possible outcomes may occur once two droplets concur to enter within the narrow channel. In the first one (panel (a)), the leading edge of a droplet undergoes a fast expansion which favors its escape towards the outlet of the constriction. In this case the droplets align horizontally within the narrow channel and no breakup occurs. In the second one (panel (b)), two droplets stretch under the effect of the viscous forces acting at the interface level, until one of them attains a critical deformation beyond which surface tension forces are not able to withstand the elongational forces, thus causing the split of the droplet in two.
It is interesting to note that, although the volume fraction of the droplets in our simulation is evidently lower than that of the experiments reported in \cite{gai2016confinement} (left rows in panel (a) and (b)), the overall deformation and breakup dynamics are still well captured by the mesoscale approach employed herewith.\\
\textcolor{black}{To gain a more quantitative insight on the deformation and breakup of the droplets, we compared the deformation history, between experiments and simulations (bottom panels of fig.\ref{FIGURA1BIS}), for the cases reported in figure. Interestingly, the agreement between the two is remarkable, since our model captures the time evolution of the deformation parameter (defined as 
$\frac{P}{2\sqrt{\pi A}}$ following Ref.\cite{gai2016confinement}, where $P$ and $A$ are the perimeter and the area of the droplet) and correctly predicts the value of its maximum deformation.}

\begin{figure}
\centering
\includegraphics[scale=0.6]{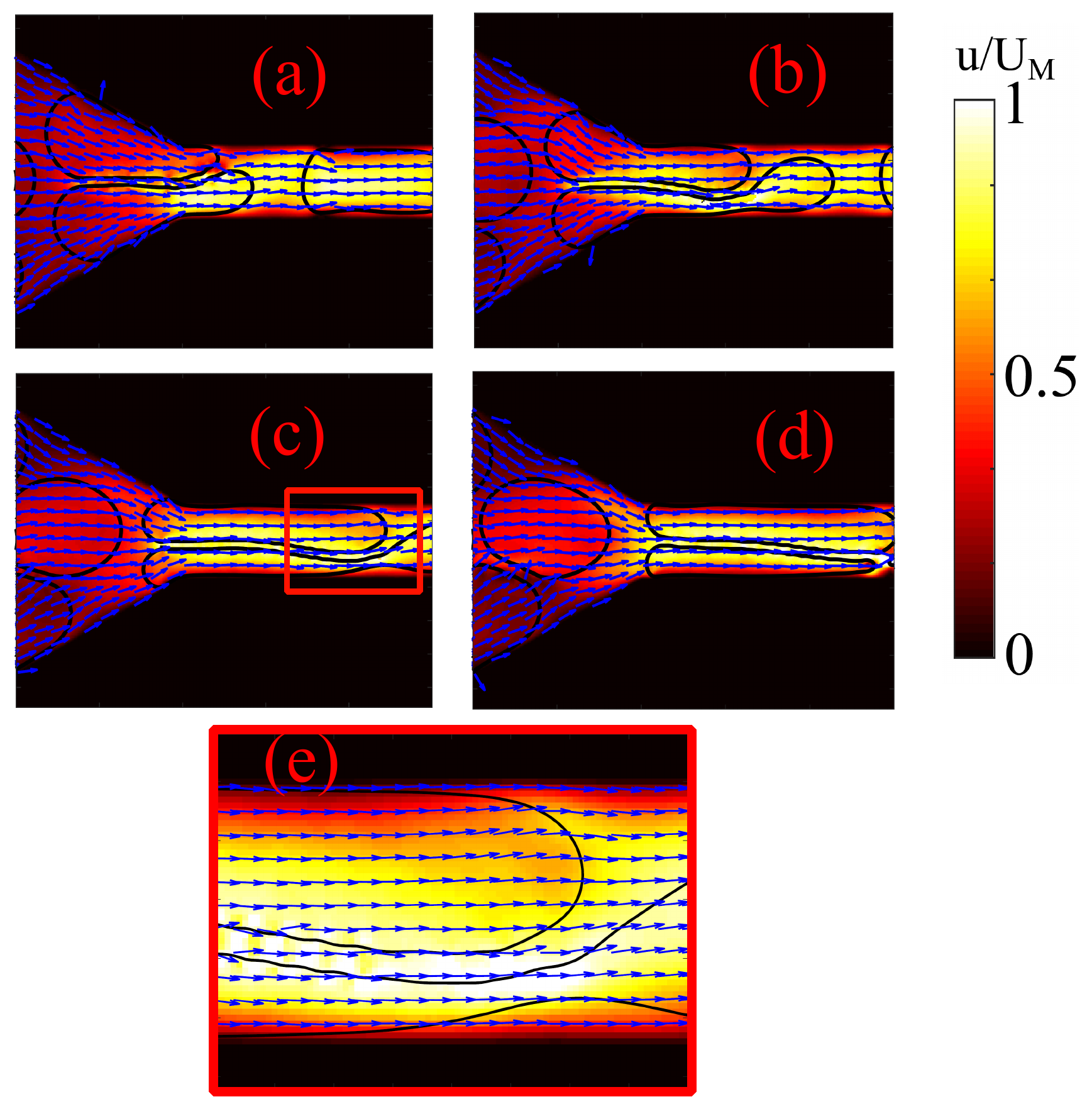}
\caption{ (a-d) Sequence of the velocity field during a breakup event at $Ca=0.043$. The color map represents the magnitude of the velocity field \textcolor{black}{divided by its maximum value in the narrow channel}, while the arrows denote the local direction of the flow field. Panel (e) shows a zoom of the area highlighted in panel (c). }
\label{fig:velfield}
\end{figure}

\subsection{\textcolor{black}{Structure of the velocity field}}

Since the LB approach grants the access to local hydrodynamic quantities not accessible to experiments, it is worthwhile inspecting the characteristics of the fluid flow developed during the breakup stages, to shed light on the complex hydrodynamic phenomena emerging in the rupture process.
\begin{figure}
\centering
\includegraphics[scale=0.85]{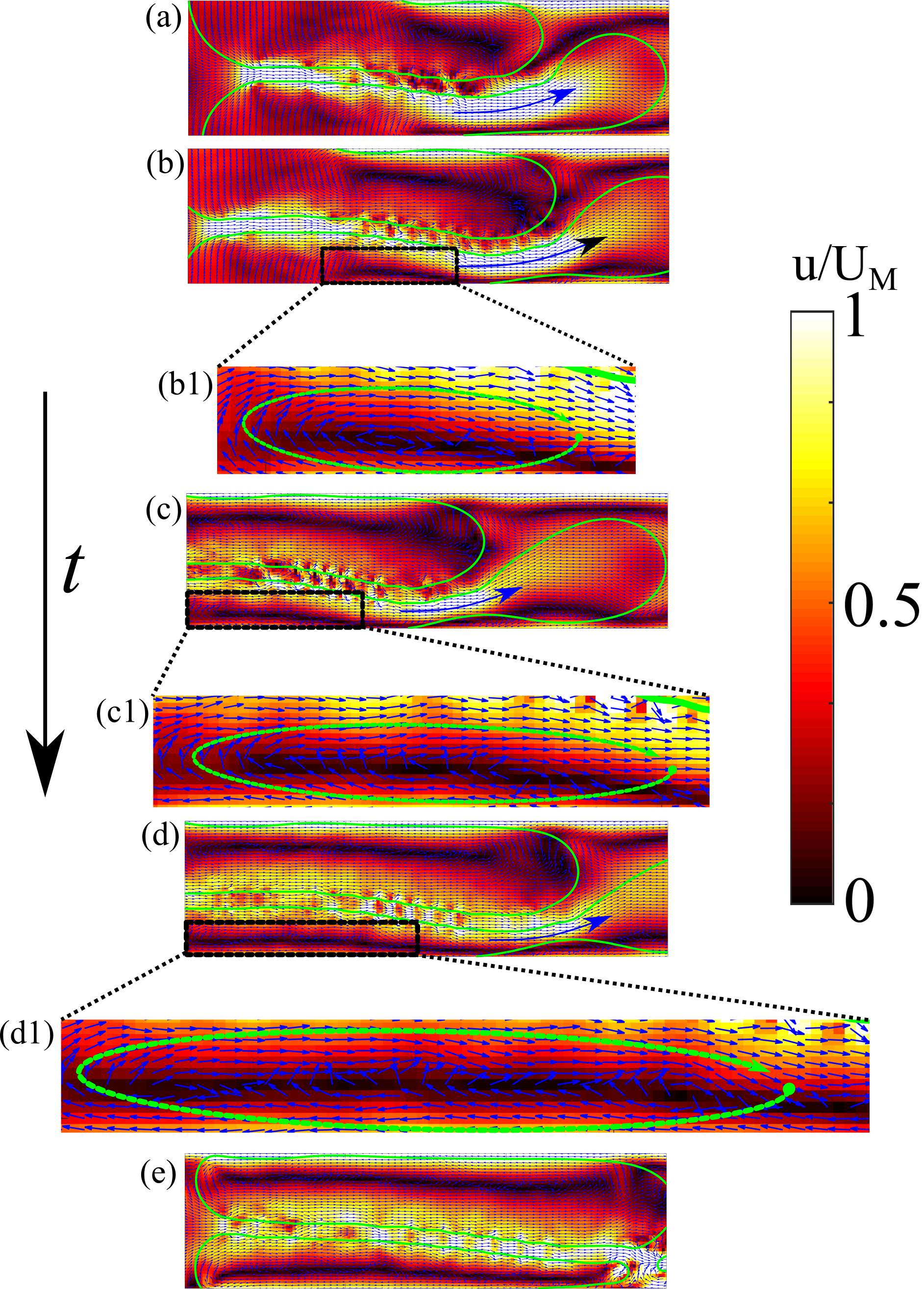}
\caption{ \textcolor{black}{(a-e) Flow field during the pinching and breakup stages, computed on a frame of reference moving with the droplets.  During the pinching stage, a recirculation forms in the upstream region, opposes to the fluid flux crossing the neck of the droplet and augments in size as the droplet moves forward and deforms. The color map denotes the magnitude of the velocity (made dimensionless since divided to the maximum value of the velocity $U_M$), while the arrows indicate the direction of the velocity field.  The arrow in the pinching region indicates the direction of the fast flow through the droplet's neck}}
\label{fig:velfield_rel}
\end{figure}
To this aim, in fig.\ref{fig:velfield} (a-d) we show the velocity field during a breakup sequence at $Ca=0.043$, where the color map represents the magnitude of the velocity and the arrows denote the local direction of the flow field.
As one can see, in an early stage (panel(a)), the leading edge of the lower droplet moves faster than its neighbor. This causes the head of the lower droplet to be quickly squeezed into the constriction by the upper one (see panel (b)). Consequently, the leading edge of the upper droplet enters the narrow channel and, due to the effect of the confinement, starts to accelerate, as evidenced by the magnitude of the flow field (see panel (c)). Afterwards, a high speed area develops at the droplet's neck (see panel (e) which shows a close-up view of the region enclosed by the rectangle in panel(c)), determining an increase of the shear between the interfaces in close contact, thus leading to the pinch-off  of the neck of the lower droplet (panel (d)).

\textcolor{black}{To further elucidate the complex fluid dynamics involved in the breakup process, we inspected the flow field during the pinching and breakup stages in a relative frame of reference moving at the average speed on each section. As shown in fig.\ref{fig:velfield_rel}, during the pinching stage a recirculation forms in the upstream region, opposes to the fluid flux crossing the neck of the droplet and increases in size as the droplet moves forward and deforms (panels b-b1 and c-c1). This means that a “fluid particle” trapped in the upstream vortex would take a time, in order to advance towards the exit of the channel, longer than the one needed by a particle flowing through the pinch. Note also that the size of the fluid recirculation increases as the droplet squeezes during its passage through the constriction, becoming comparable with the length of the upstream part of the droplet when it breaks. It is precisely this “momentum unbalance” mechanism (recurring at each breakup event) that produces the stretching of the droplet followed by its rupture.}

\textcolor{black}{To conclude this part, we would like to point out that, in each simulation, the droplets' diameter  were set to $\sim60$ lattice units, which means having a Cahn number of the order of $\sim 0.05$,  a typical value in simulating the physics of resolved, complex interfaces  (see, for instance, \cite{magaletti2013sharp}). Moreover, compressibility effects are negligible since  the maximum values of the velocity reached in our simulations are of order $10^{-2}$ (lattice units over step), delivering a Mach number of the same order. These values ensure that our approach is adequate to capture, indeed with remarkable accuracy, the correct physics observed in droplet microfluidics and that the provided fluid dynamics picture is not affected by spurious numerical artifacts.}

At this point, several questions naturally arise: What is the effect of the Capillary number, the main parameter governing the processes detailed above, on the breakup statistics?  Is there a maximum deformation over which the droplet inevitably breaks? And, if the latter exists, does it depend on the flow characteristics?
In the next sections we try to address these issues by introducing suitable observables which help to put the phenomenon under scrutiny on a more quantitative ground.

\subsection{Breakup statistics: Circularity, Breakup rates and Critical Deformation}

As introduced at the end of section IIIC, it is now of interest to investigate the behavior of the flowing emulsion from a statistical perspective, to pinpoint the effect of the governing parameters on shape deformation and breakup.

To do so, we introduce the circularity $\chi$, a dimensionless parameter assessing  the deformation of the droplets flowing within the constriction. It is defined as

\begin{equation} \label{LBM17}
\chi=\frac{4 \pi A}{P^2}
\end{equation}
and it ranges between $1$ (circular shape) and $0$ (needle-like geometry); thus monitoring its evolution allows to evaluate the departure of the droplet from its circular shape.  The observations are carried out by tracking the deformation and breakup events occurring within an Eulerian volume of control, coinciding with the narrow channel.
\textcolor{black}{Also, to gain significant statistics over the number of such events occurring in the channel, each simulation was run for $10^6$ steps.} 

\begin{figure}
\centerline{
\includegraphics[scale=0.9]{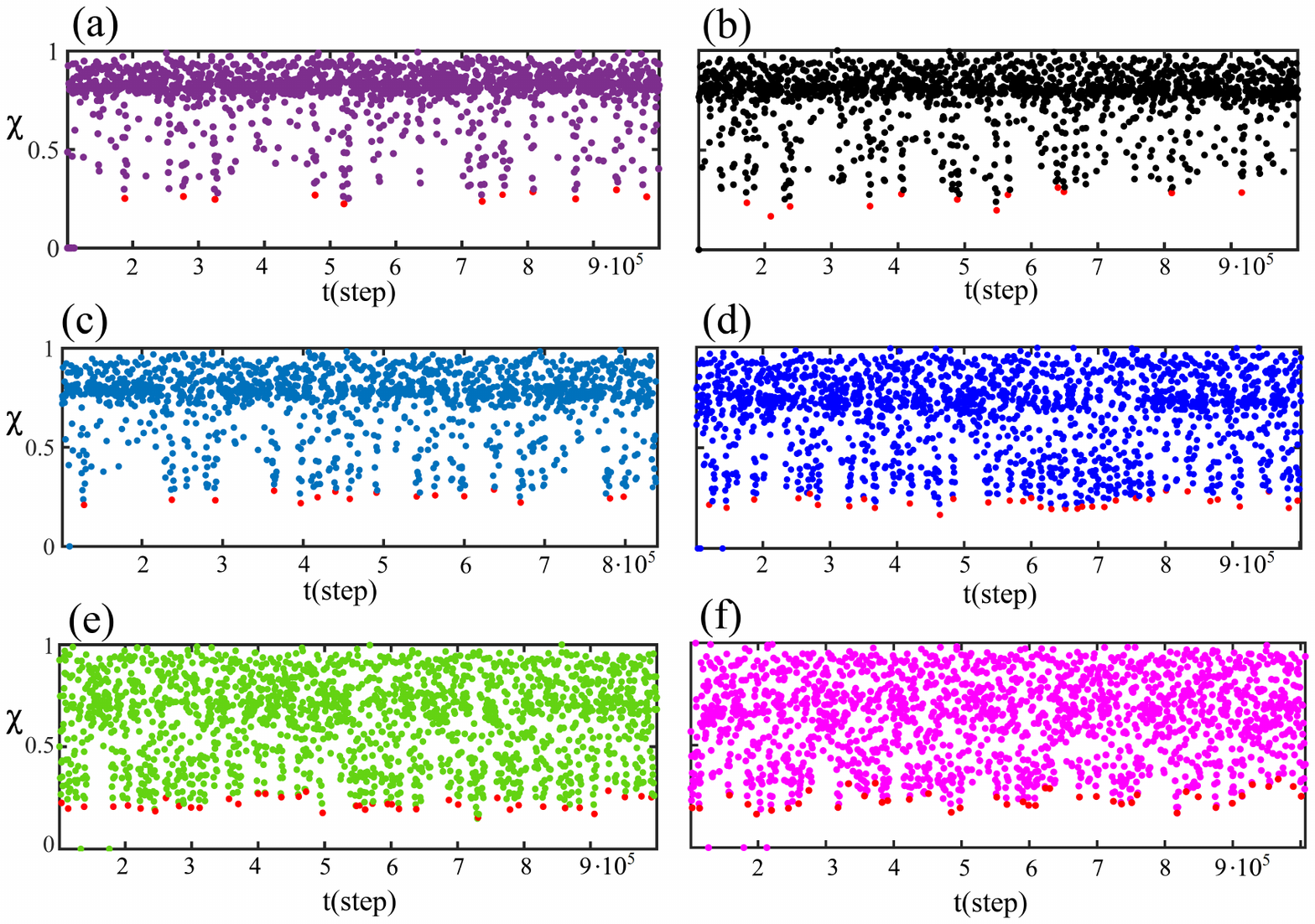}}
\caption{Circularity versus time for different Capillary numbers. (a) $Ca=0.011$. (b) $Ca=0.022$. (c) $Ca=0.029$. (d) $Ca=0.043$. (e) $Ca=0.054$. (f) $Ca=0.072$. Red spots indicate the minimum value of $\chi$, achieved before a breakup event occurs.}
\label{FIGURA2}
\end{figure}

In fig.\ref{FIGURA2} we report the plot of circularity vs time for each doublet flowing  within the constriction and for each value of the Capillary number inspected. In each plot, the red points denote the minimum value of circularity (i.e. the maximum droplet's deformation) reached just before a breakup event.
An interesting feature is that the ratio between the number of droplets which break and the total number of considered events steadily increases with $Ca$, in agreement with previous experimental data \cite{gai2016confinement}. This fact is clearly supported by the increasing number of red dots, from panel $(a)$ to box $(f)$.
Moreover, as the Capillary number increases, the distribution of points in the plots of fig.\ref{FIGURA2} appears more uniform, thus suggesting the two following observations:

i) the probability that two droplets simultaneously arrive to the constriction and compete to enter it, depends on $Ca$.

ii) when two droplets concur to enter the constriction, for low values of $Ca$ the surface tension forces prevail over the viscous (extensional) ones, thus favoring faster expansions of the droplets and preventing their breakup.   

\begin{figure}
\centerline{
\includegraphics[scale=0.8]{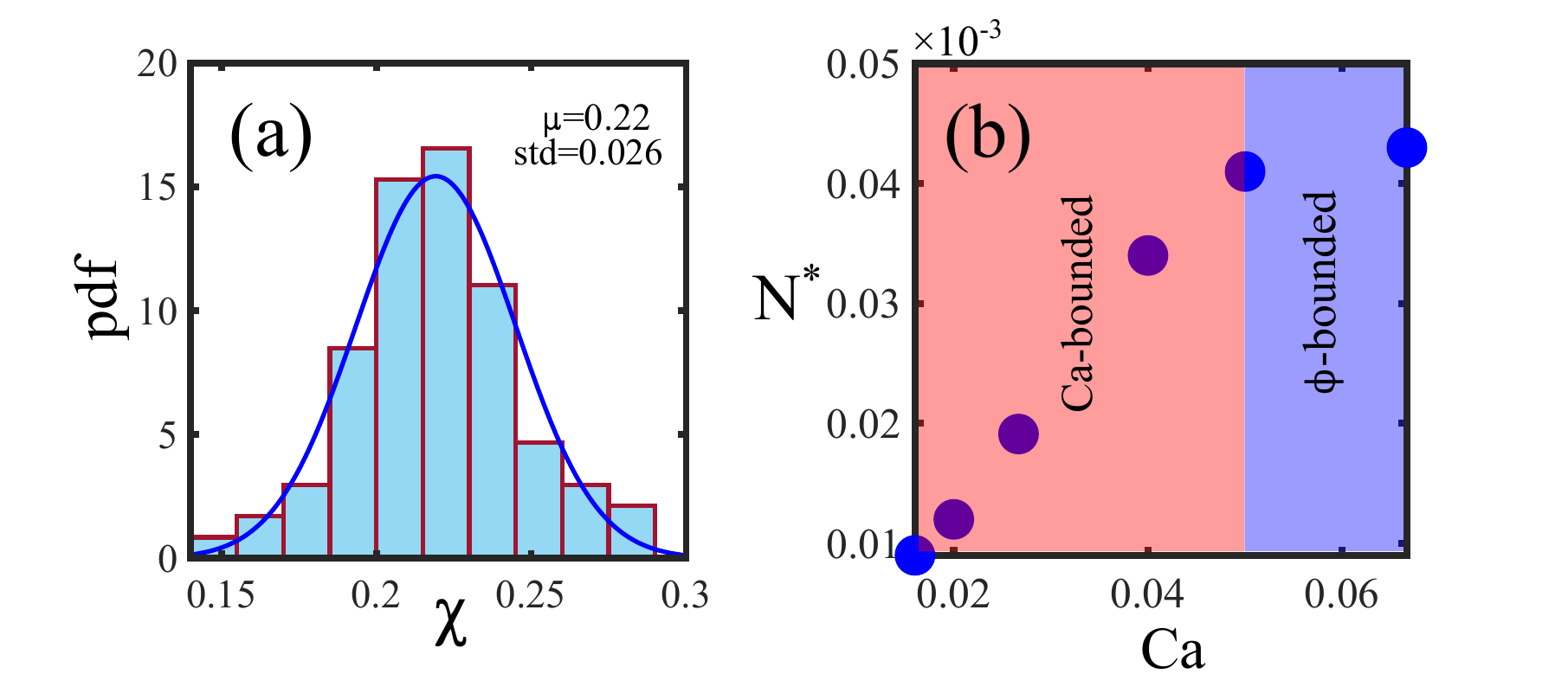}}
\caption{Panel(a): probability density function of $\chi_{min}$. Panel(b): frequency of breakup events (i.e. number of breakup events versus the total simulation time) versus $Ca$. The colored regions denote the Capillary bounded ($Ca$-bounded) and the droplets' volume fraction (($\phi$-bounded)) bounded regions.}
\label{FIGURA3}
\end{figure}

A second aspect 
is that the minimum circularity reached by the droplets before the breakup is approximately independent of the Capillary number and roughly equal to $\chi_{min}\sim 0.22$.
This observation is further confirmed in fig.\ref{FIGURA3}(a), which shows the plot of the $pdf(\chi)$ obtained by  collecting the values of $\chi_{min}$ for each breakup event. As clearly visible, the \textit{pdf} follows a Gaussian trend, centered on $\overline{\chi}_{min}\sim 0.22$ with 
a rather small standard deviation, $std\sim0.026$.
From the above, it seems reasonable to conclude that, at least for the setup under investigation and in the limit of laminar flow, the maximum deformation attained by the droplets before their breakup is independent of the flow details, being a universal characteristics of the phenomenon in play. 

On the contrary the breakup rates steadily increase as $Ca$ augments. This effect is described in fig.\ref{FIGURA3}(b), which reports the frequency of the breakup events $N^*$, defined as the ratio of the number of breakups over the number of time steps, versus $Ca$.  The breakup rate scales roughly proportional with $Ca$ in the range $0.011 \le Ca \le 0.054$, while it saturates for $Ca > 0.054$.
This fact can be explained as follows. In the first region of the plot ($Ca$-bounded), the number of breakup events is limited by the Capillary number which, as shown before, governs the deformation dynamics of the droplets concurring to enter the constriction. For low values of $Ca$, two droplets approaching the inlet of the narrow channel have a smaller probability to deform and break, since the surface tension forces dominate over the extensional actions of the viscous ones. As $Ca$ increases, the effect of the viscous forces grows accordingly and the droplets are more prone to break. By contrast, for $Ca$ larger than a critical value ($\phi$-bounded region) the number of breakup events is limited by the number of doublets competing to enter the narrow channel. 
In other words, for $Ca \ge Ca_{sat} \ (Ca_{sat} \sim 0.05)$, the number of breakup events  becomes independent of the Capillary number, attaining a saturation value which depends only on the volume fraction of the emulsion. This observation is consistent with the fact that the droplet breakup in a dense emulsion is due to droplet-droplet interactions, whose number \textcolor{black}{is expected to} considerably decrease when the volume fraction diminishes, as discussed in Ref.\cite{gai2016confinement} for a dilute emulsion.
It is thus reasonable to speculate that the critical value of the Capillary number depends on the droplet volume fraction of the emulsion. 

To conclude, it is interesting to observe that the linear relationship between the number of breakup events and the Capillary number (within the Capillary bounded region) can be deduced by calling on simple energetic arguments.
Firstly, the work per unit time done by the viscous stresses within the droplet $\dot{L}_v$ can be written as
\begin{equation} \label{LBM18}
\dot{L}_v= 2 \rho \nu \int_A  \mathcal{D} : \mathcal{D} dA \approx 2 \rho \nu \bigg( \pi \frac{U}{P} \bigg)^2 A,
\end{equation}
where $\mathcal{D}$ is the symmetric part of the velocity gradient ($\mathcal{D}=\frac{\nabla \mathbf{u} + \nabla \mathbf{u}^T}{2}$), whose order of magnitude has been estimated as $O(\mathcal{D}) \approx \pi U/P$ with $P$ and  $A$  perimeter and area of the droplet, respectively.
Moreover, it is straightforward to express $\dot{L}_v$ in terms of the circularity
\begin{equation} \label{LBM19}
\dot{L}_v \approx \frac{1}{2} \rho \nu U^2 \chi.
\end{equation}

On the other hand, the work per unit time done by the surface tension on the droplet's surface is proportional to  $\dot{L}_{st}  \approx \sigma U$.

%
By taking the ratio between  $\dot{L}_v, \dot{L}_{st}$ we can introduce a non-dimensional group $\mathcal{N}_{vst}$
\begin{equation} \label{LBM21}
\mathcal{N}_{vst}=\frac{1}{2} \frac{\rho \nu  U}{\sigma} \chi = \frac{1}{2} Ca \chi,
\end{equation}

which, interestingly, depends linearly on both $Ca$ and $\chi$.
As observed before, when a breakup occurs, the minimum circularity is independent of the Capillary number and can be considered constant while  $\mathcal{N}_{vst}$ (the ratio between extensional viscous and surface tension works per unit time) scales linearly with the Capillary number. 

\section{Conclusions}

In this work we have numerically investigated, using a lattice Boltzmann approach for multicomponent fluids augmented with disjoining near contact interactions, the deformation and breakup dynamics of droplets flowing within a tapered channel with a constriction. The built-in hydrodynamic features of such mesoscale approach allow us to study in details the fluid dynamics emerging at the interface level during the breakup phenomena. In particular, our simulations confirm the experimental observation reported in \cite{gai2016confinement}, by reproducing the peculiar behaviors observed during the deformation and breakup stages of doublets interacting within the narrow constriction.
Moreover, the introduction of a suitable parameter allows for a quantitative  assessment of the deformation state of the system. We find that i) the maximum deformation reached by the droplets before breakup is independent of $Ca$ and ii) the number of breakup events linearly depends on $Ca$ until a critical value is attained. This one is likely to have a strong dependence on the packing fraction of the emulsion.

As a perspective, the present model, properly extended, could be potentially employed in a number of biological applications, one for all hemodynamics. In this respect, an all-mesoscale approach capable of modelling the coupled evolution of plasma and red blood cells could pave the way to large scale hemodynamic simulations with unprecedented resolution and computational efficiency.

\section*{Acknowledgements}

A.M., A.T.and S.S. acknowledge funding from the European Research Council under the European Union's Horizon 2020 Framework Programme (No. FP/2014-2020) ERC Grant Agreement No.739964 (COPMAT).\\
A.M. acknowledges the CINECA Computational Grant ISCRA-C IsC83 - “SDROMOL”, id. HP10CZXK6R under the ISCRA initiative, for the availability of high performance computing resources and support.\\
M.L.R and P.P. acknowledge funding from the Italian Ministry of Education, University and Research (MIUR), in the frame of the Departments of Excellence
Initiative 2018-2022, attributed to the Department of Engineering of Roma Tre
University.

\section*{Data Availability}

The data that supports the findings of this study are available within the article



\end{document}